\documentclass[%
 reprint,
showpacs,
 amsmath,amssymb,
 aps,
pra,
floatfix,
]{revtex4-1}

\usepackage{graphicx}% Include figure files
\usepackage{dcolumn}% Align table columns on decimal point
\usepackage{bm}% bold math
\usepackage{color}
\newcommand*{\ket}[1]{\mathopen{|}#1\mathclose{\rangle}}

\newcommand{\ketbrap}[2]{\mathopen{|}#1\mathclose{\rangle}\hspace{-0.25em}\mathopen{\langle}#2\mathclose{|}}

\begin{document}

%\preprint{APS/123-QED}

\title{Dissipative Preparation of Entangled Many-Body States with Rydberg Atoms}

\author{Maryam Roghani}
\email{maryam.roghani@itp.uni-hannover.de}
% \altaffiliation[Also at ]{Institut fu\"r Theoretische Physik, Leibniz Universit\"at Hannover, Appelstraße 2, 30167 Hannover, Germany}
\author{Hendrik Weimer}
\affiliation{Institut f\"ur Theoretische Physik, Leibniz Universit\"at Hannover, Appelstrasse $2$, $30167$ Hannover, Germany
}

\date{\today}

\begin{abstract}
  We investigate a one-dimensional atomic lattice laser-driven to a
  Rydberg state, in which engineered dissipation channels lead to
  entanglement in the many-body system. In particular, we demonstrate
  the efficient generation of ground states of a frustration-free
  Hamiltonian, as well as $W$ states. We discuss the realization of
  the required coherent and dissipative terms, and we perform
  extensive numerical simulations characterizing the fidelity of the
  state preparation procedure. We identify the optimum parameters for
  high fidelity entanglement preparation and investigate the scaling
  with the size of the system.
\end{abstract}

\maketitle

\section{Introduction}

The use of controlled dissipation channels holds great promise for the
preparation of highly entangled quantum many-body states
\cite{Verstraete2009,Weimer2010}. These ideas are particularly
relevant for ultracold atoms due to the tremendous experimental
control over these systems, combined with the availability of
efficient dissipation channels in the form of optical pumping. Here,
we discuss the preparation of entangled many-body quantum states in
the context of strongly interacting Rydberg atoms.

So far, explicit state preparation protocols have been mostly
constrained to stabilizer states
\cite{Barreiro2011,Weimer2011,Carr2013a,Rao2013,Weimer2013a,Lin2013,Shankar2013},
meaning that these states can be found by minimizing the energy of a
stabilizer Hamiltonian of mutually commuting observables
\cite{Gottesman1996}. However, as shown in \cite{Verstraete2009},
efficient state preparation protocols should also exist for
non-commuting degrees of freedom in frustration-free models, where the
ground state can be found by minimizing a sum of quasilocal projection
operators.

As the physical platform to realize such dissipative quantum state
engineering, Rydberg atoms appear to be particularly suited due to
their highly tunable interaction and dissipation properties
\cite{Low2012}. Remarkably, Rydberg atoms in one-dimensional lattices
can be approximated by a frustration-free model \cite{Lesanovsky2011},
making it especially promising to study dissipative quantum state
engineering in these systems.

A crucial quantity in the investigation of strongly interacting
Rydberg gases is the blockade radius $r_b$, given by
\begin{equation}
  r_b = \sqrt[6]{\frac{C_6}{\hbar\Omega}},
\end{equation}
where $C_6$ is the van der Waals coefficient of the Rydberg
interaction and $\Omega$ is the Rabi frequency associated with the
driving of the Rydberg transition. Here, we will be interested both in
the regime where the blockade radius is comparable to the lattice
spacing and the regime where $r_b$ is larger than the system size.

This paper is organized as follows. After a brief introduction of the
atomic scheme and the theoretical approach, we discuss the
prepariation of the ground state of a frustration-free Hamiltonian. We
present our numerical results for various detuning and explore the
appropriate parameters in which the steady state overlaps with the
highly entangled Rokhsar-Kivelson state. We investigate the scaling of
the fidelity with increasing system size. Finally, we perform an
analogous analysis for the generation of entangled $W$ states, where
we find high state preparation fidelity even for relatively large
system sizes.

\section{Setup of the system}

\subsection{Hamiltonian dynamics}

We consider a one-dimensional lattice with $N$ sites. Each site is
occupied by a two level atom with states $\ket{0}$ and $\ket{1}$,
driven by a laser beam with Rabi frequency $\Omega$ and detuning
$\Delta$. The Hamiltonian for this system in the rotating wave
approximation is expressed as
\begin{equation}
\label{eq:Hamilton}
 \mathcal{H}=\Omega\sum_k^N \sigma_x^k+\Delta \sum_k^N n_k+V\sum_{m>k}^N \frac{n_m n_k}{|k-m|^6},
\end{equation}
where $n_k=\sigma_+^k \sigma_-^k$ is the Rydberg number operator in
terms of the raising and lowering operators. The atoms interact via
the van der Waals interaction with the interaction strength $V$. By
applying the unitary transformation $U=\exp[-{\rm{i}}t \sum_k^N n_k
  n_{k+1}]$ and an additional rotating wave approximation,
Eq.~(\ref{eq:Hamilton}) is rewritten as \cite{Lesanovsky2011}
\begin{equation}
\label{Hamilton1}
\mathcal{H}=E_0+\mathcal{H}_{\rm{3body}}+\mathcal{H}^{\prime},
\end{equation}
where $E_0$ is the ground state energy and $\mathcal{H}^{\prime}$ a perturbation. The three-body term $\mathcal{H}_{\rm{3body}}$ is given by
\begin{equation}
 \mathcal{H}_{\rm{3body}}=\Omega\sum_k^ N  h_k^{\dagger}h_k,
\end{equation}
i.e., it is a sum of projection operators $h_k$. These projection
operactors have the form
\begin{equation}
  h_k=\sqrt{\frac{1}{\xi^{-1}+\xi}}\,\,P_{k-1}P_{k+1}[\sigma_x^k+\xi^{-1}n_k+\xi(1-n_k)],
\end{equation}
where $P_k=1-n_k$. Within the phase diagram of one-dimensional Rydberg
gases \cite{Weimer2010a,Sela2011}, the pertubation $\mathcal{H}'$ is
minimal when the blockade radius $r_b$ is given by $r_b = 2a$ with $a$
being the lattice spacing \cite{Lesanovsky2011}. The manifold of the
region that can be appromixated by the three-body Hamiltonian
$\mathcal{H}_{\rm{3body}}$ adiabatically connects to the region
between crystalline phases at half and third filling \cite{Ji2014}.

A Hamiltonian that can be expressed as summation of the projection operators is known as frustration-free Hamiltonian with a Rokhsar-Kivelson (RK) ground state \cite{Rokhsar1988}
\begin{equation}
 |\xi\rangle =\frac{1}{\sqrt{Z_{\xi}}}\prod_k^N(1-\xi P_{k-1}\sigma_+^k P_{k+1})|00 \cdots 0\rangle,
\end{equation}
where $Z_{\xi}$ is a normalization constant. This state is a coherent
superposition of all states that has no nearest neighbour excitations
and is highly entangled. For the case $\xi=1$, the RK state can be
understood as a superoperator $\mathcal{P}$ acting on the
anti-symmetric superposition of the two atomic states,
$\prod_i|-\rangle_i=\prod_i(|0\rangle_i-|1\rangle_i)/\sqrt{2}$, where
$\mathcal{P}$ imposes a constraint prohibiting two neighbouring
excitations.

\subsection{Jump Operators}

For the dissipative preparation, we will be interested in the coupling to an external environment, which leads to the desired quantum many-body state as a steady state of the dynamics. Here, we consider an open quantum system that is described by a quantum master equation in Lindblad form
\begin{equation}
\label{eq:master}
 \frac{d}{dt}\rho(t)=-\frac{\rm{i}}{\hbar}[\mathcal{H},\rho]+\sum\limits_{k} \left(c_i\rho
    c_i^{\dagger}-\frac{1}{2}\{c_i^\dagger c_i,\rho\} \right).
\end{equation}
For the dissipative preparation of the RK state, we focus on the jump operators
\begin{equation}
\label{eq:jump}
 c_k=\sqrt{\kappa}\, P^{k-1}_g (|-\rangle \langle + |)^kP^{k+1}_g,
\end{equation}
which avoids the existence of two neighboring atoms in the excited
states, thus transfers the atoms to an antisymmetric
superpostion. In the context of Rydberg atoms, we envision three different possibilities to realize such a correlated jump operator: (i) Combining strong Rydberg interactions with electromagnetically induced transparency (EIT) \cite{Muller2009,Pritchard2010,Petrosyan2011}, it is possible to engineer effective jump operators of the desired type when the Rydberg interaction destroys the EIT feature. (ii) Realization of a rotated version of the Hamiltonian (\ref{eq:Hamilton}) using Rydberg dressing of hyperfine spin states \cite{Glaetzle2015,vanBijnen2015,Overbeck2016a}, where a $\sigma_-$ jump operator in the rotated bases corresponds to the desired $\ketbrap{-}{+}$ operator \footnote{The (rotated) projection operator $P_g$ can be incorporated by making the decay process dependent on the Rydberg-dressed interaction.} (iii) Using coherent interactions with optically pumped auxiliary atoms, allowing to realize largely arbitrary many-particle jump operators \cite{Weimer2016}.

While the jump operators of Eq.~(\ref{eq:jump}) have the RK state as a
dark state, it is not unique, as configurations involving a large
number of up spins are also dark, as there is no site on which the
constraints involving the $P_g$ operators can be fulfilled. Hence, we
add a second set of jump operators, related to optical pumping of on
spin state into the other via an intermediate state. The strength of
this second set of decay processes is chosen to be relatively weak to
the former, ensuring that the desired RK state remains an approximate
dark state of the dynamics. Interestingly, the precise form of this
second set is not very important. Even in the case (ii) outlined
above, where the decay is actually occurring in a rotated basis, we
find a very efficient decay of the undesired class of dark states. In
this case, we can integrate out the intermediate state within the
effective operator formalism \cite{Reiter2012}, see
App.~\ref{app:eff}, obtaining for these addition operators in the
rotated frame
\begin{eqnarray}
\label{eq:lindblad}
 c_{k^{\prime}}&=&\frac{\Omega'}{\sqrt{\gamma}} \rm{e}^{-\rm{i}\frac{\pi}{4} \sigma_k^y}(|0\rangle \langle 0|+|0\rangle \langle 1|) \rm{e}^{+\rm{i}\frac{\pi}{4} \sigma_k^y}, \\ \nonumber
 c_{k^{\prime \prime}}&=&\frac{\Omega'}{\sqrt{\gamma}} \rm{e}^{-\rm{i}\frac{\pi}{4} \sigma_k^y}(|0\rangle \langle 1|+|1\rangle \langle 1|) \rm{e}^{+\rm{i}\frac{\pi}{4} \sigma_k^y}.
\end{eqnarray}
In total, the jump operators $c_i$ in Eq.~(\ref{eq:master}) contain the
sets $c_k$, $c'_k$, and $c''_k$, for all lattice sites $k$.

\section{Numerical results}

\begin{figure}[t!]
\center
\includegraphics[width=\linewidth]{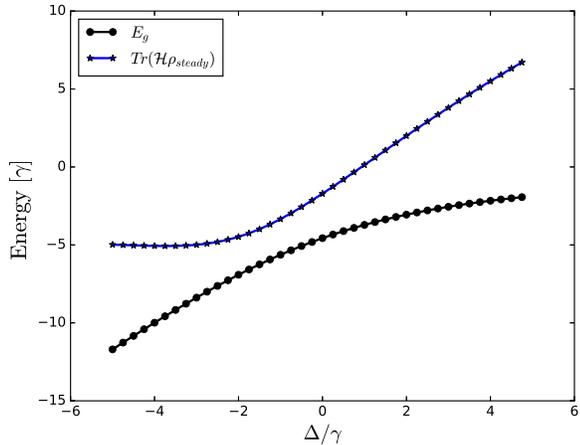}
\caption{\label{groundstate}
Comparison of the expection value of the energy in the steady state ($\star$) and the ground state energy of the  Hamiltonian $ \mathcal{H}$ ($\bullet$) for different laser detunings. Further parameters are $\Omega'^2/\gamma=0.02\kappa$, $V=100\kappa$, $\Omega=1.5\kappa$, and $N=5$. }
\end{figure}

We now perform numerical simulations of the quantum master equation
based on the wave function Monte Carlo approach \cite{Johansson2013}
to obtain the steady state $\rho_{st}$ of the
dynamics. Fig.~\ref{groundstate} compares the expectation value of the
energy in the steady state of the dynamics to the ground state energy
of the  Hamiltonian at $\xi=1$. At $\Delta=-2$, the
energies are in close agreement.

We can further explore the efficiency of our dissipative preparation
scheme by considering the fidelity $\mathcal{F}_{RK}$ of preparing the
RK state in terms of the overlap of the steady state with the RK
state,
\begin{equation}
 \mathcal{F}_{RK}=\langle \xi |\rho_{st}|\xi\rangle,
\end{equation}
where $\rho_{st}$ is the steady state density matrix from the solution
of the quantum master equation.
\begin{figure}[b!]
\center
\includegraphics[width=\linewidth]{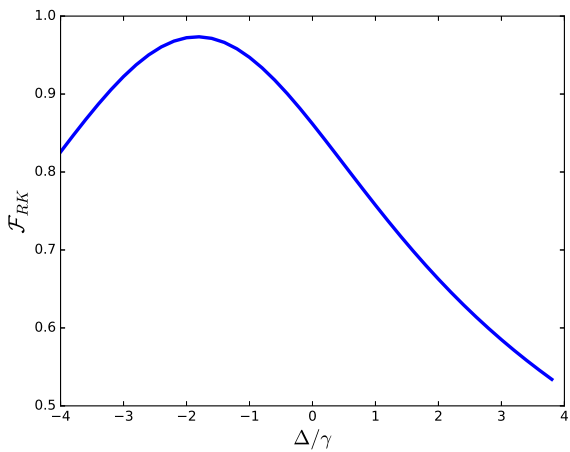}
\caption{\label{Fidelity}
Fidelity of the Rokhsar-Kivelson state, $|\xi\rangle$, as a function of detuning. Parameters are same as in Fig.~\ref{groundstate}.}
\end{figure}
Again, we find that at $\Delta=-2$, the fidelity approaches unity.

\begin{figure}[ht]
\center
\includegraphics[width=\linewidth]{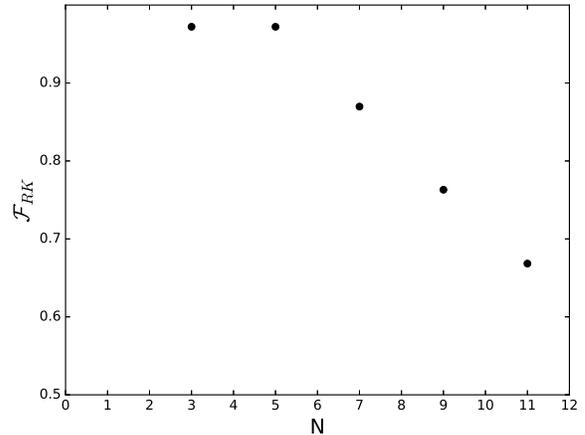}
\caption{\label{Fidelitysize}
Scaling of the fidelity of the RK state preparation with different numbers of lattice sites $N$.}
\end{figure}

Furthermore, we perform an evaluation of our preparation procedure as a
function of the system size. Generically, the fidelity is expected to
decay exponentially, as small local deviations from the RK state get
multiplied for larger system sizes, a phenomenon also known as
``orthogonality catastrophe''. Indeed, as shown in
Fig.~\ref{Fidelitysize}, we observe a decrease of the fidelity for
larger system sizes. Nevertheless, we still obtain a substantial
overlap with the RK state for system sizes as large as $N=11$. Here,
we focus on an odd number of sites, as they give slightly larger
values for the fidelity.

%\subsection{Rotated Basis}
%The jump operator defined in equation \ref{jump} is challenging because the decay occurs between the atomic eigenstates
%\begin{equation}
%\label{jumpwithout}
% c_k=\sqrt{\gamma}|\downarrow \rangle \langle \uparrow |.
%\end{equation}
%Applying rotation operator 
%\begin{equation}
% \mathcal{R}(-\rm{i}\frac{\pi}{4}\sigma_y)=exp(-\rm{i}\frac{\pi}{4}\sigma_y),
%\end{equation}
%%on Eq. \ref{jumpwithout} results in equation \ref{jump}. In other words the Hamiltonian applied in this model Eq. \ref{Hamilton} is the following equation
%\begin{equation}
%\label{notrotated}
%% {\bf{H}}=\Omega\sum_k^L \sigma_z^k+\Delta \sum_k^L  \sigma_x^k+V\sum_{m>k} \frac{|+_m +_k\rangle \langle +_m +_k|}{|r_k-r_m|^6}, 
%\end{equation}
%after applying rotation. The Lindblad operators Eq. \ref{lindblad} represented in the rotated frame. 

\section{Generation of $W$ states}

\begin{figure}[t!]
\centering
\includegraphics[width=\linewidth]{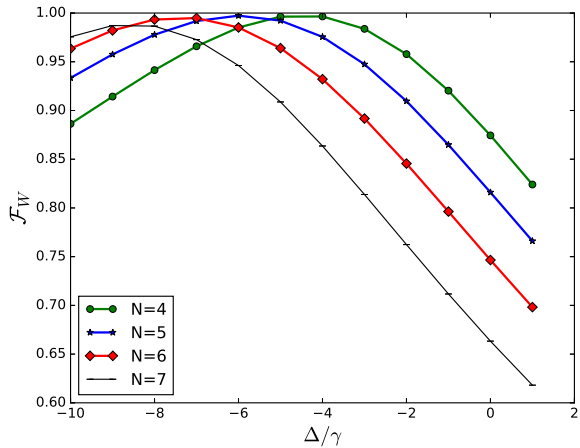}
\caption{\label{FidelityWN} Fidelity of the $W$ state prepration as a
  function of the laser detuning $\Delta$. Parameters are
  $V=10^4\kappa$, $\Omega'^2/\gamma=0.02\kappa$, $\Omega=1.5\kappa$.}
\end{figure}

Finally, we wish to discuss how the dissipation preparation procedure
can be generalized to cover other classes of entangled quantum
many-body states. Here, we consider the case where the blockade radius
is larger than the system size, $r_b\gg Na$, corresponding to a fully
blockaded ensemble \cite{Brion2007}. In such a situation, the
dynamics is mostly confined to the manifold of states containing at
most a single Rydberg excitation. In this regime, the jump operator is
defined as
\begin{eqnarray}
 c_i&=&\sqrt{\kappa}\prod_{j\neq i} \,P_g^j\, (|-\rangle \langle +|)^i \\ \nonumber
 &=& \sqrt{\kappa}P_g^1\otimes P_g^2\otimes P_g^{i-1}\otimes  (|-\rangle \langle +|)^i \otimes P_g^{i+1},
\end{eqnarray}
i.e, the quantum jump can only occur if there are no Rydberg excitations located at other sites. The dark state of interest is closely related to a $W$ state, i.e.,
\begin{equation}
 \label{wstate}
 |W\rangle=\frac{1}{\sqrt{N+1}}(|\downarrow\downarrow \cdots\downarrow\rangle-|\uparrow\downarrow\cdots\downarrow\rangle-|\downarrow\uparrow \cdots\downarrow\rangle-\cdots).
\end{equation}
As in the investigation of the dissipative preparation of the RK state, we now turn to analyzing the fidelity of the $W$ state preparation, defined as
\begin{equation}
 \mathcal{F}_{W}=\langle W |\rho_{st}|W\rangle.
\end{equation}
Figure~\ref{FidelityWN} shows the fidelity for various detunings
$\Delta$ and for different number of the atoms in the lattice. The
maximum entanglment occurs at different detunings for different number
of the atoms. Remarkably, we find that the fidelity does not
significantly decrease for larger system sizes, potentially allowing to
efficiently prepare large ensembles of entangled atoms. Performing
numerical simulations based on matrix product states \cite{Honing2012,Transchel2014,Cui2015,Werner2016} or the
variational principle for open quantum systems \cite{Weimer2015} will
allow to test the precise scaling of the fidelity with the size of the
system for much larger ensemble sizes.

\section{Summary}
We have investigated the dissipative preparation of entangled states
in one-dimensional atomic lattices in which the atoms interact via a
long-range Rydberg interaction. We find that we can efficiently
prepare moderately sized systems in a highly entangled
Rokhsar-Kivelson state, as well as larger ensembles of atoms in a $W$
state. Our results underline the strengths of the dissipative quantum
state preparation paradigm.

\section{Acknowledgments}
This work was funded by the Volkswagen Foundation and by the DFG
within SFB 1227 (DQ-mat) and SPP 1929 (GiRyd).

\appendix

\section{Jump operators in the effective operator formalism}

\label{app:eff}

One approach which eliminates the fast decaying excited states
manifold is via effective quantum jump operators \cite{Reiter2012},
which reduces the system dynamics to the non-decaying states. Here, we
will neglect the decay of the Rydberg state and focus only on the
decay of intermediate states. This procedure gives rise to an
effective Hamiltonian
\begin{equation}
\label{heff}
 \mathcal{H}_{eff}=-\frac{1}{2} V_{-}\big{[}\mathcal{H}^{-1}_{NH}+(\mathcal{H}^{-1}_{NH})^{\dagger}\big{]}V_{+}+\mathcal{H}_g,
\end{equation}
where $V_{+}/V_{-}$ are the excitation/de-excitation operators coupling to the decaying states.  $\mathcal{H}_{NH}$ is the non-Hermitian Hamiltonian of the decaying states manifold
\begin{equation}
 \mathcal{H}_{NH}=\mathcal{H}_e-\frac{\rm{i}}{2}\sum_k \mathcal{L}^{\dagger}_k\,\mathcal{L}_k,
\end{equation}
where $\mathcal{H}_e$ is the Hamiltonian in the manifold of the
decaying states and $\mathcal{L}_k$ describes the quantum jump
operators. The effective quantum jump operators are given by
\begin{equation}
 \mathcal{L}_{eff}^k=\mathcal{L}_k\mathcal{H}_{NH}^{-1}V_{+}.
\end{equation}
Applying these definition to a process that describes optical pumping from the $\ket{1}$ state into the $\ket{0}$ state via an intermediate level, we obtain
\begin{eqnarray}
\label{eq:leff}
 \mathcal{L}^0_{eff}=\frac{\rm{i}\Omega'}{\sqrt{\gamma}}(|0 \rangle\langle 0|+|0 \rangle\langle 1|),\\
 \mathcal{L}^1_{eff}=\frac{\rm{i}\Omega'}{\sqrt{\gamma}}(|0 \rangle\langle 1|+|1 \rangle\langle 1|),\nonumber
\end{eqnarray}
where $\Omega'$ is the Rabi frequency describing the driving of the
transition between $\ket{1}$ and the intermediate state and $\gamma$
is its spontaenous decay rate. After transformation into the rotating
frame, we obtain the jump operators given in Eq.~(\ref{eq:lindblad}).

\bibliographystyle{aip}
\bibliography{../../Papers/bib/bib}

\end{document}